\documentclass[11pt,a4paper]{article}
\usepackage{epsfig}

\newlength{\dinwidth}
\newlength{\dinmargin}
\def\fig#1{{Fig.~(\ref{#1})}}
\def\eq#1{{Eq.~(\ref{#1})}}
\newcommand{\Le}{\left(}
\newcommand{\Ra}{\right)}

\newcommand{\beq}{\begin{equation}}
\newcommand{\eeq}{\end{equation}}
\newcommand{\beqar}[1]{\begin{eqnarray}\label{#1}}
\newcommand{\eeqar}{\end{eqnarray}}
\begin{document}

\title {{~}\\
{\Large \bf Gas-liquid transition in the model of particles interacting at high energy}\\}
\author{ 
{~}\\
{~}\\
{\large 
S.~Bondarenko$\,\,$\thanks{Email: sergeyb@ariel.ac.il},
K.~Komoshvili$\,\,$\thanks{Email: komosh@ariel.ac.il}} 
\\[10mm]
{\it\normalsize  Ariel University Center of Samaria, Israel}\\}

\maketitle
\thispagestyle{empty}

\begin{abstract}
An application of the ideas of the inertial confinement fusion process in the
case of particles interacting 
at high energy is investigated.
A possibility of the gas-liquid transition in the gas is considered using different approaches.
In particular, a shock wave description of interactions between particles is studied and a self-similar solution of
Euler's equation is discussed. Additionally, Boltzmann equation is solved for self-consistent field (Vlasov's equation) in linear approximation for the case of a gas under external pressure and the corresponding change of Knudsen number
of the system is calculated.
\end{abstract}

%\begin{flushright}
%\vspace{-18.5cm}
%{\Large \bf DRAFT}\\
%\today
%\end{flushright}
%\thispagestyle{empty}

\newpage

\section{Introduction}

The possibility of the existence of a first-order phase transition which occurs between phases with different densities in QCD 
is a very intriguing problem for both a theoretical and experimental studying. Low energy runs at RHIC/BNL
and the future facilities FAIR at GSI and NICA in Dubna will, perhaps, clarify the experimental situation with the
signatures of the first-order phase transition in the low-medium energy interactions.
Moreover, collisions of relativistic nuclei in the RHIC and LHC experiments at very high energies led to the revelation of a new state of matter named quark gluon plasma (QGP). At initial stages of the scattering, the dense hadron gas of the
nuclei becomes almost ideal fluid, see \cite{Shyr1}. This fluid is short-living and expanding state of strongly interacting partons at very high temperature. Details of the phase transition in such interactions are not fully clear; there are arguments for the true phase transitions as well as for the fast crossover, see discussions
of the question in \cite{Shyr1,Berd,Nahrgang,Stein,Skokov,Rand}.

In our paper, exploring some "toy" model, we will try to 
understand, at least qualitatively, the process of the gas-liquid transition which can take place in a system of 
interacting relativistic particles. We assume, that the proposed "toy" model can be applied for the case of high-energy 
scattering as well as for the case of scattering at low and medium energies, providing in both cases an initial conditions for further evolution of the system.
The process of nuclei-nuclei scattering was widely investigated in the frameworks of different
thermodynamics approaches. The statistical mechanics techniques were widely used for the description of
systems of two interacting nuclei as well as for the description of 
the data on the multiplicities of produced particles, see \cite{Land1,Multip}.
The possibility of  applications of Boltzmann
equation for the description of the scattering process was investigated as well, see \cite{BolVlas}.
Nevertheless, the attempts to describe the nuclei in the framework of Boltzmann equation with the help 
of global distribution functions of nuclei do not look quite satisfactorily. 
Nuclei
scattering at high energy is highly non-equilibrium process, see for example \cite{Berd,Step}. 
The scattering system can not be in a global equilibrium 
state, instead, only some local spots of the scattering area can be locally equilibrated, see \cite{Tor} and also
 \cite{Berd,Skokov} for more details. Therefore, in our model we consider only local spots of the equilibrated matter which, in turn, are influenced by the matter outside of the spots. 

\begin{figure}[t]
\begin{center}
\psfig{file=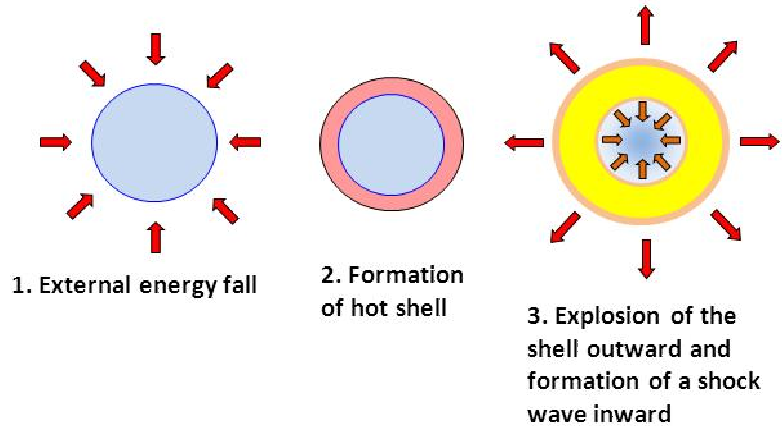,width=110mm} 
\end{center} 
\caption{\it Inertial confinement in the thermonuclear plasma fusion.}
\label{InCon}
\end{figure}

The assumption about the locally equilibrated spots of the hot matter inside the scattering region we combine with the results of papers \cite{Land1} and \cite{Term}. The main idea of \cite{Term} is that non-equilibrium process
can be considered as an equilibrium one but in an external field. It leads to the  separation of the matter
locally in the scattering area into two groups. For the first group of the matter, we use an equilibrium description whereas the 
second group is considered as a source of the external field. The difference between these two groups can be understood 
from \cite{Land1}, see QCD degrees of freedom separation in \cite{Eff1}. Following the ideas of \cite{Land1},  
we assume that the hot drops of the hadron matter with very high density are created in the scattering region
at the very first moment of the interactions. Approximately half of the total initial energy of the interacting partons
is going there. The particles inside the drops we can call as internal ("slow") or stopped particles. The initial energy of the "slow"
particles in each spot is spend to heating of the dense drop. Other partons  are outside of the drops of stopped particles. These outside particles 
we call as external or "fast", important that they remain relativistic. 
Therefore, the influence of the "fast" particles on the process is realized in further compression of the hot drops  and
can be accounted by introduction of some external field acting on the internal particles.
The advantage of this approach is clear. We reduce the non-equilibrium process to the equilibrium one 
and, therefore, we can apply a wide variety of equilibrium kinematic approaches
separately to each local spot.
The whole process of the dense drop compression, therefore, is similar to the effect of inertial confinement in plasma physics, see \cite{InConf}. The radiation pressure of the external area (hot shell) of the target
is used in order to create a requested 
density of the matter, see Fig.\ref{InCon}. Another advantage of the approach is that the relativistic dynamics will 
be important only for the external field description, the "slow" particles undergo a non-relativistic 
transport process.

 The "toy" model of the problem is a gas consisting of the colliding disks which are characterized by rapidity variable.
The value of the rapidity variable separate the "fast" from the "slow" particles. 
Namely, the rapidity of the external particles is much larger than the rapidity of the initially stopped particles. 
The use of this variable is justified not only by the picture of QCD high energy interactions governed by the Pomeron amplitude,
see \cite{Eff1} and \cite{bfkl}, but
in general, by the clear energy dependence of the possible phase transition. 
Choosing the rapidity of the stopped particles as zero, we assume that
the potential of the external field created by the external particles depends on the rapidity of
these particles which is proportional to the total energy of the scattering process.  

Further, 
in Section 2, we develop our model of the gas and
introduce a pairwise potential between the particles in two dimensional space. The external field in the model depends
on the rapidity variable and it is the only characteristics of the external particles (relativistic dynamics) in this framework. For our purposes, the form of the potential is unimportant and some simple one is chosen. The following subsections of this section are dedicated to the modified Van der Vaals equation of the state of the gas,
see also \cite{Skokov}. Section 3 is dedicated to the thermodynamical properties of the gas. In particular, basing on Enskog approach, see \cite{Enskog,Prakash}, we discuss the transport characteristics of the gas. In Section 4  the shock wave description of the process of the interaction of the external and  internal particles is given.
There we describe the influence of the external particles on the internal ones in the form of shock wave and solve Euler's equations  following
mainly by the results of paper \cite{Shock}.
In Section 5, we generalize the problem and consider Boltzmann equation in  Vlasov's approximation for the case of a collisonless plasma in three dimensions and describe an equilibrium state of the system\footnote{An 
investigation of the deviation from the equilibrium state is considered in the Appendix B of the paper}. In each  Section the question addressed to the issue of the gas-liquid transition is also discussed. We investigate the conditions and signs of this transition for each of the proposed frameworks.
The last section of the manuscript is dedicated to the discussion of the obtained results; there the conclusion of the paper is given.

%%%%%%%%%%%%%%%%%%%%%%%%%%%%%%%

\section{Equation of the state of the gas}

\subsection{The potential of a pairwise interactions}

Our "toy" model is a system of two relativistically colliding two dimensional disk clouds. Due to the Lorentz
contraction, important already at not so high energies, see \cite{Cser}, we restrict our consideration by two dimensional model\footnote{The model can be easily extended for three dimensions, but in this case we will 
loose an applicability of the calculations for the high-energy interactions. We plan to investigate similar 3-dimensional
model ever later. Nevertheless, 3-dimensional Boltzmann equation is considered in Section 5.}.
Therefore, similarly to high energy scattering of two nuclei, we assume that at very early stages of the collision two dimensional hot spots of stopped particles are created. For the formulation of the equation of state of these spots we need to determine the pairwise interaction potential
between the particles\footnote{We limit ourselves only by pairwise interactions.}, see Appendix A. 
In our task, the following potential consisting of two parts is considered:
\beq\label{Poten}
U(b)\,\,=\,\int\,f(b-x)\,V_{1}(b)\,f(x)\,d^2\,x\,+\,V_{2}(b)\,.
\eeq
Here $V_{1}(b)$ is the usual non-relativistic pairwise potential arising between the particles at the mutual distance $b$ inside the drop:
\beq
V_{1}(b_{12})\,=\,-A/b\,+\,B/b^{\alpha}
\eeq
with $\alpha\,>\,1$ (in our further calculations we take $\alpha\,=\,2$) and $A,B$ are some constants. 
The function $f(x)$ is the distribution function of the particles which we take equal to 
\beq\label{DistF}
f(x)\,=\,\frac{e^{-x^{2}/R^2}}{\pi\,R^2}\,,
\eeq 
$R$ being a radial size, see \cite{Levin} for justification of that choose.
The potential energy term $\,V_{2}(b)\,$ , absent in \eq{TEn}, describes the potential energy of a 
particle inside of the spot in the field of the "fast" particles outside the spot.
This term accounts the influence of the "fast" particles on the dynamics of "slow" particles\footnote{The value of $R$ is somehow arbitrary, it can be considered as the radius of proton for example, or as a characteristic length of hot spot inside the collision region.} and this is the only source of the relativistic dynamics in the problem. 
Therefore, similarly to high-energy relativistic model of \cite{Levin}, this external potential field $\,V_{2}(b)\,$ is assumed to be
depended only on the position (impact parameter) $b$ of the particle inside of the region, and rapidity $Y\,=\,\ln\Le\frac{s}{s_0}\Ra$ of the
outside particles in respect to the "slow" particles; here $s$ is the squared total energy of the "fast" particles in c.m.f. of the "slow" process. 
This field creates additional pressure on the internal particles and could
be considered as an analogue of the inertial confinement effect in plasma physics. For simplicity
we choose the form for this term 
similar to what we have in the phenomenological Pomeron approach (see \cite{Levin}) :
\beq\label{Radius}
V_{2}(b)\,=\,-\,\frac{C}{2\,\pi\,R^2}\,e^{Y}\,\theta\,(R^2\,-\,b^2\,)\,
\eeq 
where $\theta$ is the step function and $C$ is a constant. 
With this potential, \eq{Poten} for the total potential energy acquires the following form:
\beq\label{Poten1}
U(b)=\frac{e^{-b^2/2R^2}}{2\pi\,R^2}V_{1}(b)\,+\,V_{2}(b)=
\frac{e^{-b^2/2R^2}}{2\pi\,R^2}\Le -A/b+B/b^{2}\Ra\,-\,
\frac{C}{2\,\pi\,R^2}\,e^{Y}\,\theta\,(R^2\,-\,b^2\,)\,.
\eeq
Considering the integral \eq{Integr} with the potential \eq{Poten1}
we see that one can approximate this integral by the following expression
\beq
J_{2}\,=\,\int\,d^{2}\,{b}\,\Le\,
\exp\left(-\beta\,\,U_{12}\,\right)\,-\,1\,\Ra\,\approx\,-\int_{0}^{b_0}\,d^{2}\,b\,-
\,\beta\,\int_{b_0}^{\infty}\,U_{12}\,\,d^{2}\,b\,,
\eeq
where $b_0$ is the value of $b$ where
\beq\label{Cond}
\,U_{12}(b_0)\,\simeq\,0\,.
\eeq
Thereby, the value of $b_0$ in \eq{Integr} we find as the roots of the following simple equation:
\beq
-\,\frac{A}{b}\,+\,\frac{B}{b^2}\,-\,C\,e^{Y}\,\,=\,0
\eeq
whose solution in the limit of large $Y$ (high energy limit) is
\beq\label{BValue}
b_{0}(Y)\,\approx\,\,e^{-Y/2}\,\Le\,\sqrt{\frac{B}{C}}\,-\,\frac{A}{2C}\,e^{-Y/2}\,\Ra\,=\,
\,\tilde{b}_{0}(Y)\,e^{-Y/2}\,.
\eeq
Clearly, at very large values of final rapidity $Y$, we obtain that $b_{0}\,\ll\,B/A$, where $B/A$ is the inter-particle distance when the external pressure is absent.
The reason of this decrease of the inter-particle distance is the presence of the additional pressure introduced in the model.
The important observation of these simple calculations is that we observe here a decrease of so-called
Knudsen number, see \cite{Klim}, which is 
\beq\label{Knud1}
\frac{b_{0}\,A}{B}\,\ll\,1\,.
\eeq
The decrease of the value of Knudsen number, in turn, indicates the transition of the gas to the liquid state under the influence of the external pressure, see again 
\cite{Klim}. This observation, in fact, does not depend on the number of dimensions of the spot. For the case of radial 
symmetry the result will be the same as in the case of three dimensional spot.

\subsection{Equation of the state of the gas}

Now, we calculate the integral \eq{Integr}:
\beq
J_{2}\,=\,\int\,d^{2}\,b\,\Le\,
\exp\left(-\beta\,\,U_{12}\,\right)\,-\,1\,\Ra\,\approx\,-\int_{0}^{b_0}\,d^{2}\,b\,-\,\beta\,\int_{b_0}^{\infty}\,
U(b)\,\,d^{2}\,b\,=\,
\eeq
$$
-\pi\,b_{0}^{2}+\frac{A\beta}{R^{2}}\int_{b_0}^{\infty}\,e^{-b^2/2R^2}\,db-
\,\frac{B\beta}{2\,R^{2}}\int_{b_0}^{\infty}\,\frac{e^{-b^2/2R^2}}{\,b^2}\,db^2
+\frac{C\beta\,e^{Y}}{R^2}
\int_{b_0}^{\infty}\,\theta\,(R^2\,-\,b^2)\,b\,db\,.
$$
Performing the integration and keeping in the integrals only the leading terms at large $Y$ ($b_0(Y)\,\ll\,1$) 
we obtain:
\beq
J_{2}\,\approx\,-\pi\,b_{0}^{2}\,+\,\beta\,\frac{A}{R}\sqrt{\frac{\pi}{2}}\Le 1-\sqrt{\frac{2}{\pi}}\,
\frac{\tilde{b}_{0}}{R}e^{-Y/2}\Ra\,+\,\beta\,\frac{B}{2R^2}\,\Le\,\gamma\,-\,\frac{\tilde{b}_{0}^2}{2R^2}e^{-Y}\Ra\,+\,
\eeq
$$
\,+\,\frac{\beta\,e^{Y}\,C\,}{2\,R^2}\,\Le\,R^2\,-\,b_{0}^{2}(Y)\,\Ra\,.
$$
Using shorter notations, we write the answer as
\beq
J_{2}\,\approx\,-\pi\,b_{0}^{2}\,+\,\beta\,\Le\,\tilde{A}(Y)\,+\,\tilde{B}(Y)\,+\,\tilde{C}(Y)\,e^{Y}\,\Ra\,.
\eeq
Therefore, the state equation of the gas in our case has the following form:
\beq
P\,=\,\frac{NT}{S}\,-\,\frac{N^2\,T}{S^2}\,\Le\,
-\pi\,b_{0}^{2}(Y)\,+\,\beta\,\Le\,\tilde{A}(Y)\,+\,\tilde{B}(Y)\,+\,\tilde{C}(Y)\,e^{Y}\,\Ra\,\Ra\,,
\eeq
or, using the shorter notations again, we have:
\beq
P\,=\,
\frac{NT}{S}\,-\,\frac{N^2\,T}{S^2}\,\Le\,-\,\pi\,b_{0}^{2}(Y)\,+\,\beta\,a(Y)\,\Ra\,.
\eeq 
Introducing the gas density variable 
\beq
\rho\,=\,\lim_{N,S\rightarrow\,\infty}\,\frac{N}{S}\,,
\eeq
we rewrite the state equation in the following form:
\beq\label{StateEq}
\Le\,P\,+\,\rho^2\,a(Y)\,\Ra\,\Le\,1\,-\,\rho\,\pi\,b_{0}^{2}(Y)\,\Ra\,=\,\rho\,T\,.
\eeq
This equation of state of the gas is the modified Van der Vaals equation with the coefficients which are depend on the
rapidity of the process. The following critical parameters of the gas-liquid transition are found from this  equation:
\beq\label{CrV3}
\rho_{cr}\,=\,\frac{1}{3\pi\,\tilde{b}_{0}^{2}}\,e^{Y}\,,\,\,\,\,P_{cr}\,=\,\frac{\tilde{C}}{27\,\pi\,\,\tilde{b}_{0}^{2}}\,e^{5Y/2}\,,\,\,\,\,
T_{cr}\,=\,\frac{8\,\tilde{C}}{27\,\pi\,\,\tilde{b}_{0}^{2}}\,e^{5Y/2}\,.
\eeq

An approach with a modified Van der Vaals equation in two spatial dimensions was also considered 
in \cite{Skokov} for the description of the hadron-quark phase transition in heavy ion collision. We briefly will discuss this approach in the Conclusion of the paper.

\section{Transport properties of the gas }

The equation of state, \eq{StateEq}, we rewrite it in the following form:
\beq\label{StateEq1}
P\,=\,Z(Y,T)\,\rho\,T\,,
\eeq
where
\beq\label{ComprFac}
\,Z(Y,T)\,=\,\frac{1}{1\,-\,\rho\,\pi\,b_{0}^{2}(Y)}\,-\,\beta\,\rho\,a(Y)\,
\eeq
This equation of state, \eq{StateEq1}, we further rewrite in the  form valid in high energy limit: 
\beq\label{StateEq11} 
P\,=\,\rho\,T\,\Le\,1\,+\,\eta\,Z(\eta)\,\Ra\,,
\eeq
where 
\beq
Z(\eta)\,=\,\frac{1}{1\,-\,\eta}\,.
\eeq
In this equation, the compressibility factor $Z(\eta)$ from  \eq{StateEq11} coincides with 
Enskog factor in \eq{StateEq1}. Therefore, the transport properties of the gas, which are characterizing by
shear viscosity $\zeta$, bulk viscosity $\xi$ and heat conductivity $\kappa$ are detrmined by the 
calculations of \cite{Enskog}:
\beq\label{TrProp1}
\zeta\,=\,\zeta_0\,\eta\,\Le\,\frac{1}{\eta\,Z(\eta)}\,+\,1\,+\,0.87\,\eta\,Z(\eta)\,\Ra\,,
\eeq
\beq\label{TrProp2}
\xi\,=\,\xi_0\,\eta\,\Le\,1.25\,\eta\,Z(\eta)\,\Ra\,,
\eeq
\beq\label{TrProp3}
\kappa\,=\,\kappa_0\,\eta\,\Le\,\frac{1}{\eta\,Z(\eta)}\,+\,\frac{3}{2}\,+\,0.87\,\eta\,Z(\eta)\,\Ra\,.
\eeq
The Enskog approximation, therefore, allows to find the relative changes of the transport coefficients
of the system under the pressure
whereas the initial values of these coefficients are given, see also calculations of \cite{Prakash}.
In the limit of high energy, whereas $\eta\,\ll\,1$, keeping only the leading in $\eta$ terms, we obtain:
\beq\label{TrProp11}
\zeta\,\approx\,\zeta_0\,,
\eeq
\beq\label{TrProp22}
\xi\,=\,1.25\,\xi_0\,\eta^2\,Z(\eta)\,\approx\,\,1.25\,\xi_0\,e^{-2Y}\,\Le\,\pi\,\tilde{b}_{0}^{2}(Y)\,\rho\,\Ra^2\,\Le\,
1\,+\,e^{-Y}\,\pi\,\tilde{b}_{0}^{2}(Y)\,\rho\,\Ra\,,
\eeq
\beq\label{TrProp33}
\kappa\,=\,\kappa_0\,\Le\,1\,+\,\frac{1}{2}\,\eta\,\Ra\,.
\eeq
We see, that the bulk viscosity $\xi$ is small in the limit of large final rapidity $Y$ compared to the case
when $Y=0$ for our gas.

The  results obtained for the transport coefficients we can consider
from the point of view of the results for the quark-gluon plasma models. Calculations of \cite{Rat}  demonstrated that the $\xi\,/\zeta\,$
ratio has a maximum at some critical temperature $T_{c}$. Considering the same ratio constructed from \eq{TrProp11}-\eq{TrProp22} values we obtain that this ratio is equal to
\beq
\frac{\xi}{\zeta}\,=\,\frac{\,1.25\,\xi_0\,e^{-2Y}\,
\Le\,\pi\,\tilde{b}_{0}^{2}(Y)\,\rho\,\Ra^2\,}{\zeta_0}\,
\eeq
and that it is small. 
Moreover, this ratio has a maximum at $\rho\,=\,\rho_{cr}$ and at critical temperature $T_{c}$\footnote{The critical temperatures in \cite{Rat} and here have different meanings, but both of them define higher temperature comparing to initial one.} corresponding to the critical parameters of the Van der Vaals equation \eq{StateEq}:
\beq
\frac{\xi}{\zeta}\,=\,\frac{1.25\,\xi_0\,}{9\,\zeta_0}\,,
\eeq
which is similar to the properties of the same ratio in the quark-gluon plasma model of \cite{Rat}. Thereby our simple model demonstrates properties which are similar to the properties of 
much more complicated models of quark-gluon plasma. This result means that in the Enskog approximation, the bulk viscosity is proportional to the density of the gas and the maximum values of the density and bulk viscosity are achieved at the liquid state of the system. Further, when the state of incompressible liquid is achieved due the evolution, the density and the bulk viscosity of the liquid remain constant.

\section{Shock wave point of view}

In this section, we proceed following mainly by the results of paper \cite{Shock}. Our flow 
of the compressed interacting particles can be described by the following
Euler's equations for the radial velocity, mass density
\beq\label{Sh1}
\partial_{t}\sigma\,+\,\partial_{r}(\sigma\,u)\,+\,\frac{1}{r}\,\rho\,u\,\,=\,0\,,
\eeq
\beq\label{Sh2}
\partial_{t}\,u\,+\,u\,\partial_{r}\,u\,+\,\frac{1}{\sigma}\,\partial_{r}\,\sigma\,=\,0\,,
\eeq
and, assuming an absence of dissipative processes for our gas, for the entropy
\beq\label{Sh3}
\partial_{t}\,s\,+\,u\,\partial_{r}\,s\,=\,0\,.
\eeq
Here $u$ is the radial velocity of the flow, $\sigma\,=\,\rho\,m\,$ is the mass density of the gas and $s$ is the entropy of the system:
\beq\label{Sh4}
s\,=\,\ln\frac{P}{\sigma^2\,\exp(C(\sigma))}\,+\,s_0\,
\eeq
with
\beq
C(\sigma)\,=\,-\,\ln\Le\,1-\sigma\,\pi\,b_{0}^{2}(Y)/m\Ra^2\,,
\eeq
see \cite{Shock}. There are three partial differential equations for the main fields which characterize the flow of
interacting particles under the external pressure. Solution of this system is considered in the next subsection.

\subsection{Shock wave description of the process: self-similar solution}
 
The process of the gas compression can be considered as 
the process of the propagating convergent radial shock wave described by the scaling variable
\beq
\xi\,=\,\frac{r}{r_{shock}}\,=\,\frac{r}{A\,(t_{0}\,-\,t)^{\alpha}} 
\eeq
where $A$ and $\alpha$ are some constants to be determined later. 
An implosion of the shock wave occurs at time $t\,=\,t_{0}$ which corresponds to $\xi\,=\,\infty\,$ and the front of the shock wave is described by the 
radius
\beq\label{Shock}
r_{shock}=A\,(t_{0}\,-\,t)^{\alpha}\,.
\eeq
The fields of interests are assumed to have the following form:
\beq\label{Dens}
\sigma\,=\,\sigma_{0}\,G(\xi)\,,
\eeq
\beq\label{Vel}
u\,=\,\frac{\alpha\,r}{t\,-\,t_{0}}\,V(\xi)\,,
\eeq
\beq\label{Pres}
P\,=\,\frac{\alpha^2\,r^2}{2\,(t\,-\,t_{0})^2}\,\sigma_{0}\,G(\xi)\,W(\xi)\,,
\eeq
with functions $G(\xi)\,V(\xi)\,W(\xi)\,$ to be found from the equations \eq{Sh1}\,-\,\eq{Sh4}
which we rewrite as
\beq\label{DEq1}
\frac{d\,V}{d\,\ln \xi}\,+\,\Le\,V\,-\,1\Ra\,\frac{d\,\ln G}{d\,\ln \xi}\,+\,2\,V\,=\,0\,,
\eeq
\beq\label{DEq2}
\Le V-1\Ra\frac{d\,V}{d\,\ln \xi}+\frac{W}{2}\frac{d\,\ln G}{d\,\ln \xi}+\frac{1}{2}
\frac{d\,W}{d\,\ln \xi}+W-V\Le\,\frac{1}{\alpha}-V\Ra\,=\,0
\eeq
\beq\label{DEq3}
\frac{d\,W}{d\,\ln \xi}-\frac{d\,\ln G}{d\,\ln \xi}-\frac{d\,C}{d\,G}\,\frac{d\,G}{d\,\ln \xi}+
\frac{2}{\alpha}\,\frac{1-\alpha\,V}{1-V}\,=\,0\,.
\eeq
The system of the differential equations \eq{DEq1}\,-\,\eq{DEq3}
can be linearized by expansion of the functions  $V,W$ around $V(\xi\rightarrow\infty)\rightarrow\,0$ and
$W(\xi\rightarrow\infty)\rightarrow\,0$ that corresponds to $t\,\rightarrow\,t_{0}$ time in \eq{Vel}\,-\,\eq{Pres}
, see more details in \cite{Shock}.
The solutions of the linearized equations can be easily found:
\beq
V\,\simeq\,\frac{K_{V}}{\xi^{1/\alpha}}\,,
\eeq
\beq
W\,\simeq\,\frac{K_{W}}{\xi^{2/\alpha}}\,,
\eeq
with constants $K_{V}$ and $K_{W}$ must be found from the full equations.
Concerning the $G$ function from \eq{Dens} we note, that this function remains finite
at $\xi\rightarrow\infty$:
\beq
\frac{d\,\ln G}{d\,\ln \xi}\,\rightarrow\,0
\eeq
when $V(\xi\rightarrow\infty)\rightarrow\,0$ and
$W(\xi\rightarrow\infty)\rightarrow\,0$.

In order to solve the equations \eq{DEq1}\,-\,\eq{DEq3}, we need initial values of our functions
$V(\xi=1),Z(\xi=1),G(\xi=1)$ which could be determined from the matching equations on the discontinuities of the
flow:
\beq
\sigma_{1}\,u_{1}\,=\,\sigma_{2}\,u_{2}\,,
\eeq
\beq
P_{1}\,+\,\sigma_{1}\,u^{2}_{1}\,=\,P_{2}\,+\,\sigma_{2}\,u^{2}_{2}\,,
\eeq
\beq
U_{1}\,+\,\frac{P_{1}}{\rho_{1}}\,+\,\frac{m\,u^{2}_{1}}{2}\,=\,U_{2}\,+\,\frac{P_{2}}{\rho_{2}}\,+\,\frac{m\,u^{2}_{2}}{2}\,.
\eeq
with subscript $1$ for the quantities before the shock and  subscript $2$ for the quantities after.
Solutions of these equations in the limit $P_{2}\gg\,P_{1}$ were found in \cite{Shock} and they are the following:
\beq\label{Match1}
\sigma_{2}\,=\,\sigma_{1}\,\Le\,1\,+\,\frac{2}{Z(\eta_{2})}\,\Ra\,,
\eeq 
\beq
P_{2}\,=\,\frac{2\,\sigma_{1}\,u_{1}^{2}}{2\,+\,Z(\eta_{2})}\,,
\eeq
\beq
u_{2}\,-\,u_{1}\,=\,-\,\frac{2\,u_{1}}{2\,+\,Z(\eta_{2})}\,,
\eeq
\beq
\,u_{1}\,=\,-\,\dot{r}_{shock}\,=\,\frac{\alpha\,r_{shock}}{t\,-\,t_{0}}\,.
\eeq
Additional equation which relates the densities of the flow is \eq{Dens}:
\beq
\sigma_{2}\,=\,\sigma_{0}\,G(1)\,=\,\,\sigma_{1}\,G(1)\,
\eeq
that together with \eq{Match1} gives:
\beq
G(1)\,=\,1\,+\,\frac{2}{Z(\rho_{0}\,\pi\,b_{0}^{2}(Y)\,G(1))}\,=\,1\,+\,2\,\Le\,1\,-\,\rho_{0}\,\pi\,b_{0}^{2}(Y)\,G(1)\,\Ra\,.
\eeq
Therefore, we obtain for $G(1)$:
\beq
G(1)\,=\,\frac{3}{1\,+\,\rho_{0}\,\pi\,b_{0}^{2}(Y)\,}\,
\eeq
that in the high energy limit $Y\,\gg\,1$ gives $\rho_{0}\,\pi\,b_{0}^{2}(Y)\,\ll\,1$ and
\beq
G(1)\,\approx\,3
\eeq
for any value of $\rho_{0}$\,.
The functions $V(1)$ and $W(1)$ could be found as well and they have the following form:
\beq
V(1)\,=\,1\,-\,\frac{1}{G(1)}
\eeq
and
\beq
W(1)\,=\,2\,\frac{G(1)\,-\,1}{G(1)^2}\,.
\eeq
The numerical values of $\alpha\,=\,0.8\,$ and $G(\infty)\,=\,4.6$ are also calculated in \cite{Shock}.

\subsection{Parameters of the solution and fluid state of the gas}
 
In order to calculate the parameter $A$ from the \eq{Shock} we use the condition
on the front of shock wave at initial time $t\,=\,0$ :
\beq
r(t\,=\,0)\,=\,A\,t_{0}^{\alpha}\,=\,R\,
\eeq
with $R$ from \eq{DistF}.
So far we obtain
\beq
A\,=\,\frac{R}{\,t_{0}^{\alpha}\,}\,.
\eeq
On the other hand, the velocity of the shock wave at the initial moment may be found from \eq{Vel}:
\beq
c_{sh}\,=\,V(1)\,\frac{\alpha\,R}{t_{0}\,}\,,
\eeq
that gives
\beq
t_{0}\,=\,\alpha\,V(1)\,\frac{R}{c_{sh}}\,.
\eeq
Finally we obtain:
\beq
A\,=\,\frac{R}{\,t_{0}^{\alpha}}\,\,=\,\Le\,\frac{c_{sh}}{\alpha\,V(1)}^{\alpha}\Ra\,R^{1\,-\,\alpha}\,.
\eeq
Now we consider the pressure, \eq{Pres},  achieved in the system of the interest with known solution for $G,V,W$ functions:
\beq\label{Pres1}
P\,=\,\frac{\alpha^2\,r^2}{2\,(t\,-\,t_{0})^2}\,\sigma_{0}\,G(\xi)\,W(\xi)\,=\,
\frac{\alpha^{2}\,\sigma_{0}\,G(\xi)\,K_{W}}{2}\,A^{2/\alpha}\,r^{2\,-\,2/\alpha}\,.
\eeq
Before the compression of the gas the state equation of the gas is the usual one:
\beq
T_{0}\,=\,\frac{P_{0}}{\rho_{0}}\,,
\eeq
whereas at the time of the compression \eq{StateEq1} holds:
\beq
T(\xi)\,=\,\,\frac{P(\xi)}{\rho(\xi)\,Z(\eta)}\,.
\eeq
Therefore, with the pressure from \eq{Pres1} we obtain for the temperature ratio:
\beq
\frac{T(\xi)}{T_{0}}\,=\,\frac{\alpha^{2}\,\sigma_{0}\,\rho_{0}}{2\,\rho(\xi)\,P_{0}\,Z(\eta)}\,
G(\xi)\,K_{W}\,A^{2/\alpha}\,r^{2\,-\,2/\alpha}\,.
\eeq
Simplifying the expression we obtain finally
\beq
T(\xi)\,=\,\frac{m\,c_{sh}^{2}}{2\,\,Z(\eta)\,V(1)^{2}}\,
\,K_{W}\,\Le\,\frac{R}{r}\Ra^{2/\alpha\,-\,2}\,.
\eeq
The maximum temperature at the center of the area is achieved when $\xi\,\rightarrow\,\infty$ at $r\,\rightarrow\,0$.
Taking this limit and changing  $r$ in the expression by the minimal possible distance \eq{BValue}, that smooth out the divergence, we obtain the maximum temperature
achieved in the system in comparison to the initial temperature $T_{0}\,=\,P_{0}/\rho_{0}$:
\beq\label{TRatio1}
T(\xi=\infty)\,=\,\,e^{\Delta\,Y}\,\,\frac{m\,c_{sh}^{2}}{2\,\,Z(\rho_{0}\,\pi\,b_{0}^{2}(Y)\,G(\infty)\,)\,V(1)^{2}}\,
\,K_{W}\,\Le\,\frac{R}{\tilde{b}_{0}(Y)}\Ra^{2/\alpha\,-\,2}\,
\eeq
with
\beq
\Delta\,=\,\frac{1\,-\,\alpha}{\alpha}\,\approx\,0.25.
\eeq
The maximum temperature achieved can be also compared 
with the temperature $T_{0}\,=\,T(\xi\,=\,1)$ on the edge of the shock wave. In this case we have:
\beq\label{TRatio}
\frac{T(\xi)}{T(1)}\,=\,\frac{Z(\eta(1))}{G(1)\,Z(\eta(\xi))}\,\Le\,\frac{R}{r}\Ra^{2/\alpha\,-\,2}\,
\eeq
and again smoothing out the divergence we obtain:
\beq\label{TRatio2}
\frac{T(\infty)}{T(1)}\,=\,e^{\Delta\,Y}\,
\frac{\,Z(\rho_{0}\,\pi\,b_{0}^{2}(Y)\,G(1))\,}{G(1)\,Z(\rho_{0}\,\pi\,b_{0}^{2}(Y)\,G(\infty)\,)}\,\Le\,\frac{R}{\tilde{b}_{0}(Y)}\Ra^{2/\alpha\,-\,2}\,.
\eeq
The resulting expressions \eq{TRatio1}, \eq{TRatio2} are interesting from the following point of view.
Taking in \eq{TRatio2} $T(\infty)$ as a critical temperature of the gas-liquid transition \eq{CrV3} we obtain:
\beq
\frac{T(\infty)}{T(1)}\,\propto\,\frac{T_{cr}}{T(Y=0)}\,\approx\,e^{5Y/2}
\eeq
that together with \eq{TRatio2} gives:
\beq\label{Knud}
\frac{R}{\tilde{b}_{0}(Y)}\,\approx\,e^{\Delta^{'}\,Y}\,>\,1
\eeq
where
\beq
\Delta^{'}\,=\,\frac{7\alpha\,-\,2}{4\,-\,4\alpha}\,.
\eeq
As mentioned above, it is well known, see \cite{Klim} for example, that when the characteristic length of the system, which is $R$ in our case, 
is larger than the average distance between particles, which is $\tilde{b}_{0}(Y)$ approximately, then
we can consider a fluid flow instead initial gas state\footnote{The ratio opposite to \eq{Knud} ratio is an analog of Knudsen number in hydrodynamics.}. Thereby we obtain an important result: in our model an indication of the gas-liquid transition
is given by the dynamical change of the Knudsen number of the problem. The possibility of calculation of the dynamical change of the Knudsen number is
considered in the next section of the manuscript.

\section{Boltzmann equation for the self-consistent field}

In this section, we generalize the model and consider a hot spot of the gas of the
charged particles interacting  in three dimensional space\footnote{We formulate the problem in 3 dimensions because of importance of longitudinal dimension in real
high energy interactions. The radial symmetry of the problem in this formulation is preserved as well and in the following we assume that all vectors of interest are radial. Therefore the vector notations
will be use only when it will be need in.} under the pressure of the external particles. The Vlasov approach
to the Boltzmann equation is valid when dissipative processes are negligible, in this case a local equilibrium is achieved.
We assume that this is the case for each separate drop and, thereby, we 
consider the following Vlasov equation for one-particle distribution function:
\beq\label{Bolz}
\frac{\partial\,f(r,p,t)}{\partial\,t}\,+\,v\,\frac{\partial\,f(r,p,t)}{\partial\,r}\,+\,
F(r)\,\frac{\partial\,f(r,p,t)}{\partial\,p}\,=\,0\,
\eeq
where we adopted the radial symmetry of the problem, and where
\beq\label{Force}
F(r)\,=\,\frac{\partial\,V_{2}(r)}{\partial\,r}\,+\,q\,E\,=\,
\frac{\partial\,V_{2}(r)}{\partial\,r}\,-\,q\,\frac{\partial\,\phi_{0}(r)}{\partial\,r}\,
\eeq
is a force which consists of two terms arose from the self-consistent electrical potential and some external potential. 
The electric  field
$E\,=\,-\,grad\,\phi_{0}(r)$ in \eq{Force} is a self-consistent electric field created by
our charged particles. This field is assumed to be weak enough in order to justify
the linear approximation for the distribution function. We assume also that the magnetic field is small 
and we neglect it in our calculations\footnote{More realistic calculations will require the magnetic field inclusion for sure.}. The Vlasov's system of equations include 
Maxwell's equations for the electric field:
\beq\label{Max}
rot\,E\,=\,0\,,\,\,\,\,div\,E\,=\,4\,\pi\,q\,n\,\int\,f(r,p,t)\,d^{3}\,p\,,
\eeq
where $n$ is a particle's density.

\subsection{Linear approximation: equilibrium state}

We solve the Boltzmann equation \eq{Bolz} representing the distribution function as
an equilibrium one with a small linear correction\footnote{The corrections to the equilibrium distribution function and electric field are calculated in the Appendix B.}:
\beq\label{Bolz1}
f(r,p,t)\,=\,f_{0}(r,p)\,+\,f_{1}(r,p,t)\,,\,\,\,f_{1}\,\ll\,f_{0}\,.
\eeq
The electric field has the same functional form as well:
\beq
E\,=\,E^{0}\,+\,E^{1}\,.
\eeq
We are looking for an equilibrium\footnote{In general, this state is quasi-equilibrium,
the external pressure acts during a finite period of time. Nevertheless we assume that this time is 
long enough and we could consider this state approximately as an equilibrium one.} state of a hot spot in the presence of an external pressure. 
The Boltzmann equation which describes the equilibrium state of the hot spot has the following form:  
\beq\label{Bolz2}
\,v\,\frac{\partial\,f_{0}(r,p)}{\partial\,r}\,+\,
F(r)\,\frac{\partial\,f_{0}(r,p)}{\partial\,p}\,=\,0\,,
\eeq
and the solution of this equation is the Boltzmann-Maxwell distribution function:
\beq\label{Equil}
f_{0}(r,p)\,=\,\frac{1}{\Le\,2\,\pi\,k_{B}\,m\,T\,\Ra^{3/2}}\,
e^{-\frac{p^2}{2mk_{B}T}\,+\,\frac{V_{2}(r)}{k_{B}T}\,-\,\frac{q\,\phi_{0}(r)}{k_{B}T}}\,.
\eeq
The Maxwell equation for the electric field \eq{Max}, therefore, reduces to non-homogeneous
Helmholtz equation which has the following form in the limit of high temperatures (large kinetic energies
of the particles in comparison to the potential energy):
\beq
-\,\Delta\phi_{0}(r)\,=\,4\,\pi\,q\,n\,\Le\,1\,+\,\frac{V_{2}(r)}{k_{B}T}\,-\,
\frac{q\,\phi_{0}(r)}{k_{B}T}\,\Ra\,
\eeq
or
\beq\label{Max1}
\,\Delta\,\phi_{0}(r)\,-\,\,4\,\pi\,q^2\,n\,\frac{\phi_{0}(r)}{k_{B}T}\,=
\,-\,4\,\pi\,q\,n\,\Le\,1\,+\,\frac{V_{2}(r)}{k_{B}T}\,\Ra\,.
\eeq
We could rewrite it in the following form:
\beq
\,\Delta\,\phi_{0}(r)\,-\,\frac{\phi_{0}(r)}{r_{D}^2}\,=
\,-\,Q_{0}\,-\,\frac{V_{2}(r)}{q\,r_{D}^{2}}\,,
\eeq
with $Q_{0}\,=\,4\,\pi\,q\,n\,$ as the charge density and 
\beq\label{Debye}
r_{D}^{2}\,=\,\frac{k_{B}\,T}{4\,\pi\,q^2\,n\,}
\eeq 
as the Debye length. The solution of this equation is the sum of the solutions of homogeneous and
non-homogeneous Helmholtz equations:
\beq\label{Poten11}
\phi_{0}(r)\,=\,Q_{0}\,r_{D}^{2}\,+\,C_{0}\,\frac{e^{-\,r\,/\,r_{D}}}{r}\,+\,
\int\,\frac{\,r^{'2}\,dr^{'}}{q\,r_{D}^{2}}\frac{V_{2}(r^{'})\,e^{-|\,r\,-\,r^{'}\,|/r_{D}}}
{\,|r\,-\,r^{'}\,|}\,,
\eeq 
where $C_{0}$ is some constant determined by the boundary conditions of the problem. In respect that the first term in r.h.s. of \eq{Poten11} is a constant, we obtain the
final expression for the potential:
\beq\label{Poten2}
\phi_{0}(r)\,=\,C_{0}\,\frac{e^{-\,r\,/\,r_{D}}}{r}\,+\,
\int\,\frac{\,r^{'2}\,dr^{'}}{q\,r_{D}^{2}}\frac{V_{2}(r^{'})\,e^{-|\,r\,-\,r^{'}\,|/r_{D}}}
{\,|r\,-\,r^{'}\,|}\,.
\eeq 
When the external field is absent we obtain usual expression:
\beq\label{Poten3}
\phi_{0}(r)\,=\,C_{0}\,\frac{e^{-\,r\,/\,r_{0D}}}{r}\,
\eeq 
where we introduced the sign $\,r_{0D}$ for the Debye length 
in the case when the external pressure is absent. 

In order to estimate the $r$ dependence of \eq{Poten2} potential we will use the 
following simple form of the external field potential:
\beq\label{ExtPot}
V_{2}(r)\,=\,V(Y)\,\theta\,(r_{D}\,-\,r)\,
\eeq
with the coefficient $V(Y)$ which depends only on the total rapidity of the process. The substitution
of \eq{ExtPot} into \eq{Poten2} gives:
\beq\label{Poten4}
\phi_{0}(r)\,=\,C_{0}\,\frac{e^{-\,r\,/\,r_{D}}}{r}\,+\,\frac{V(Y)}{q\,r_{D}^{2}}
\int_{0}^{r_{D}}\,\,r^{'2}\,dr^{'}\frac{\,e^{-|\,r\,-\,r^{'}\,|/r_{D}}}
{\,|r\,-\,r^{'}\,|}\,.
\eeq 
In the case of the potential \eq{ExtPot} the solutions for the electric potential \eq{Poten4} must be considered
separately in the two different regions: 
\begin{enumerate}
\item Region where $r\,<\,r_{D}$. In this case we have:
\beq\label{Poten5}
\phi_{0}(r)\,=\,C_{0}\,\frac{e^{-\,r\,/\,r_{D}}}{r}\,+\,\frac{V(Y)}{q\,r_{D}^{2}}\Le\,
\int_{0}^{r}\,\,r^{'2}\,dr^{'}\frac{\,e^{-(\,r\,-\,r^{'}\,)/r_{D}}}
{\,(r\,-\,r^{'}\,)}\,+\,\int_{r}^{r_{D}}\,\,r^{'2}\,dr^{'}\frac{\,e^{-(\,r^{'}\,-\,r\,)/r_{D}}}
{\,(\,r^{'}\,-\,r\,)}\,\Ra.
\eeq 
Keeping only leading in $r/r_{D}$ terms in the integration, we obtain: 
\begin{eqnarray}\label{Poten6}
\phi_{0}(r)\,& = &\,C_{0}\,\frac{e^{-\,r\,/\,r_{D}}}{r}\,
+\,\frac{V(Y)}{q\,r_{D}^{2}}\Le\,r^2-2r\,r_{D}\Le e^{-1+r/r_{D}}-e^{-r/r_{D}}\Ra + \right.\\
&\,& + \left.
r_{D}\Le\,2r_{D}-e^{-1+r/r_{D}}(2r_{D}-r)-e^{-r/r_{D}}(r+r_{D})\Ra\Ra\,.\nonumber
\end{eqnarray}

\item Region where $r\,>\,r_{D}$. For these values of $r$, we obtain:
\beq\label{Poten7}
\phi_{0}(r)\,=\,C_{0}\,\frac{e^{-\,r\,/\,r_{D}}}{r}\,+\,\frac{V(Y)}{q\,r_{D}^{2}}\,
\int_{0}^{r_{D}}\,\,r^{'2}\,dr^{'}\frac{\,e^{-(\,r\,-\,r^{'}\,)/r_{D}}}
{\,(r\,-\,r^{'}\,)}\,.
\eeq 
Integrating this expression and again keeping only leading $r_{D}/r$ terms
we obtain:
\beq\label{Poten8}
\phi_{0}(r)\,=\,C_{0}\,\frac{e^{-\,r\,/\,r_{D}}}{r}\,+\,\frac{V(Y)}{q}\,
e^{-\,r\,/\,r_{D}}\Le\,B_{0}\,+\,B_{1}\,\frac{r}{r_{D}}\,+\,\,B_{2}\,\frac{r^2}{r_{D}^2}\,\Ra\,
\eeq
where $B_{0}\,,B_{1}\,,B_{2}\,$ are some constants.
\end{enumerate}

\subsection{Debye length changes}

The value of the Debye length, \eq{Debye}, depends on the particle density $n$ which may vary 
depending on the external conditions, i.e. the particle density  of hot spot is not a constant. Therefore, we consider another parameter, called plasma parameter, which is defined as 
\beq\label{Deb1}
\mu\,=\,\frac{1}{n\,\int^{r_{D}(n)}\,d^{3}x\,\int\,d^{3}p\,f_{0}(r,p)}\,\ll\,1\,,
\eeq
and which we require to stay the same in the presence and absence of the external field $V_{2}$. 
From \eq{Deb1} we see that in the case of the external pressure the Debye radius is smaller than
in the case of the absence of the pressure:
\beq\label{Deb}
(r_{D}(Y))_{V_{2}\,\neq\,0}\,<\,(r_{D})_{V_{2}\,=\,0}\,.
\eeq 
Using   \eq{Poten3}, \eq{Poten6} we can estimate this effect of the change in the Debye length.
First of all, consider the inverse of the parameter \eq{Deb1} in the absence of the external field:
\beq\label{Plas}
1\,\ll\,\frac{1}{\mu}\,=\,4\pi\,n\,\int^{r_{D}(n)}\,r^2\,dr\,\int\,d^{3}p\,\,f_{0}(r,p)\,
\eeq
where
\beq
f_{0}(r,p)\,=\,\frac{1}{\Le\,2\,\pi\,k_{B}\,T\,\Ra^{3/2}}\,
e^{-\frac{p^2}{2mk_{B}T}\,-\,\frac{q\,\phi_{0}(r)}{k_{B}T}}\,.
\eeq
with $\phi_{0}(r)$ from \eq{Poten3}:
\beq
\phi_{0}(r)\,=\,C_{0}\,\frac{e^{-\,r\,/\,r_{0D}}}{r}\,.
\eeq
Simple integration gives:
\beq\label{Plas1}
1\,\ll\,\frac{1}{\mu}\,\approx\,4\pi\,n\,\int^{r_{D}(n)}\,r^2\,dr\,\Le\,1\,-\,\frac{q\,\phi_{0}(r)\,}{k_{B}\,T\,}
\Ra\,=\,
\,N\,\Le\,1\,-\,\frac{C_{1}\,q\,}{k_{B}\,T\,r_{0D}}\Ra\,,
\eeq
where $N\,=\,\frac{4}{3}\pi\,r_{D}^{3}\,n\,$ is the number of particles in the spot and $C_{1}$
is a positive constant. In the presence of the external field we have instead \eq{Poten3} the
expression \eq{Poten2} with the potential given by \eq{Poten6}. Therefore, in the leading in $r/r_{D}$
order, we will obtain:
\beq\label{Plas2}
1\,\ll\,\frac{1}{\mu}\,\approx\,4\pi\,n\,\int^{r_{D}(n)}\,r^2\,dr\,\Le\,1\,-\,\frac{q\,\phi_{0}(r)\,}{k_{B}\,T\,}
\,+\,\frac{V(Y)}{k_{B}\,T\,}\Le\,A_{0}\,-\,\frac{r}{r_{D}}\,A_{1}\,+\,
 \frac{r^{2}}{r^{2}_{D}}\,A_{2}\,\Ra\,\Ra\, 
\eeq
where $A_{0}\,,A_{1}\,,A_{2}\,$ are some positive constants.
Integrating \eq{Plas2} we obtain:
\beq\label{Plas3}
1\,\ll\,\frac{1}{\mu}\,=\,\,N\,\Le\,1\,-\,\frac{C_{1}\,q\,}{k_{B}\,T\,r_{D}}\,+\,
\frac{V(Y)\,C_{2}}{k_{B}\,T\,}
\Ra\,
\eeq
with some positive constant $C_{2}$\footnote{It is important that $C_{2}$ is a positive. In our calculations we obtained
$C_{2}\,=\,\frac{41}{10\,e}\,-\,\frac{3}{4}$.}. We require that both expressions \eq{Plas1} and \eq{Plas3} 
are equal and this gives for the Debye's length changes:
\beq
\,\frac{r_{0D}\,-\,r_{D}}{r_{0D}\,r_{D}}\,=\,C_{2}\,\frac{V(Y)}{q}\,
\eeq
or
\beq
\,r_{D}\,=\,\frac{r_{0D}}{\,1\,+\,\,C_{2}\,V(Y)\,r_{0D}\,/\,q}\,
\eeq
with $C_{2}$ as some positive constant. As it was underlined in previous Section, see also \cite{Klim}, the 
hydrodynamic description of the process is possible when the characteristic 
length of the system, in our case $r_{D}$, begins to decrease, i.e. Knudsen number of the system begins to decrease as well. We determine the Knudsen number as 
\beq\label{HydIneq}
Kn\,=\,\frac{r_{D}}{r_{0D}}\,=\,\frac{1}{\,1\,+\,\,C_{2}\,V(Y)\,r_{0D}\,/\,q}\,<\,1\,,
\eeq
and we indeed obtain that this number decreases when the external pressure is applied.

\begin{figure}[t]
\begin{tabular}{c c}
\psfig{file=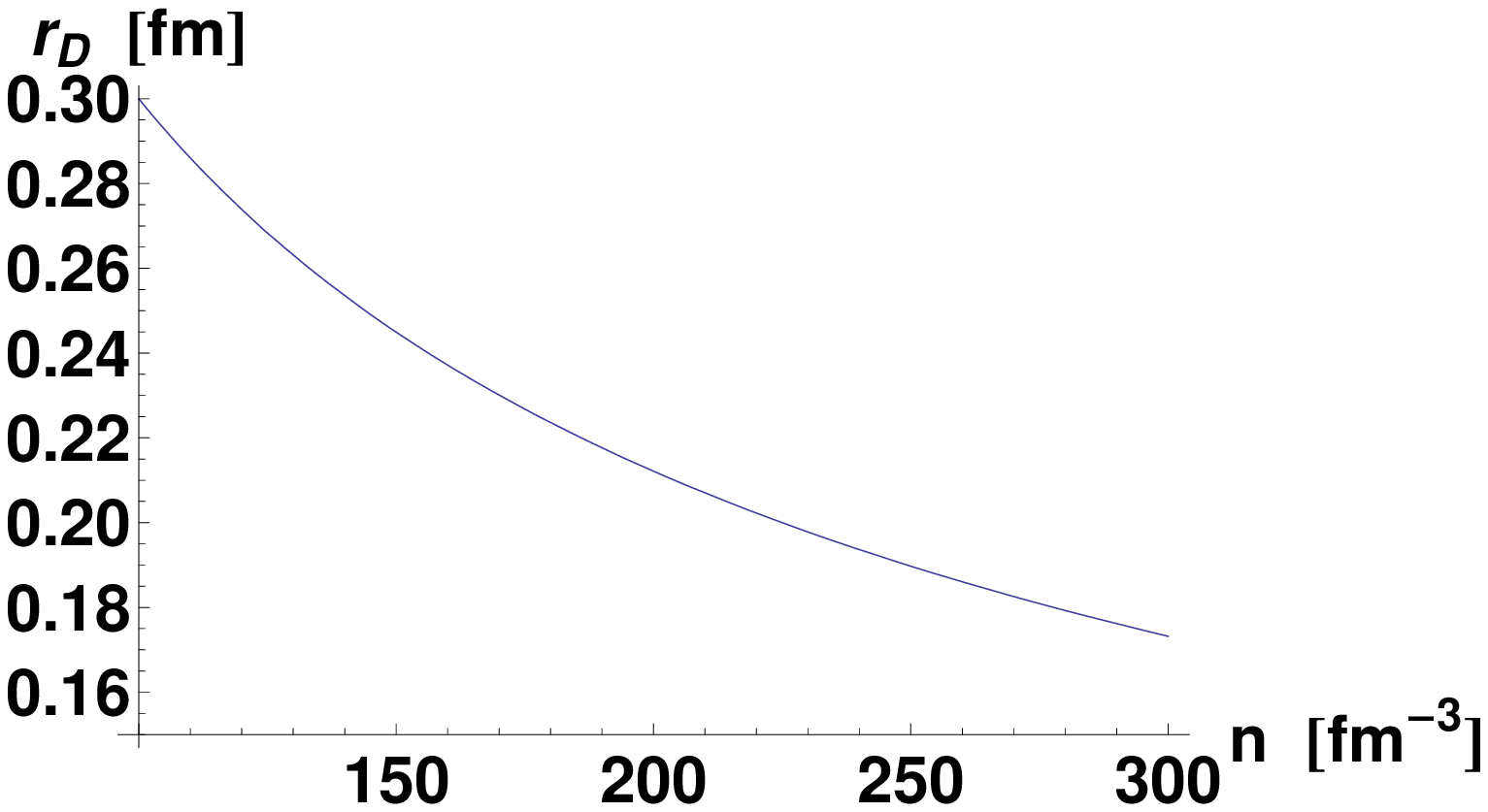,width=90mm} & 
\psfig{file=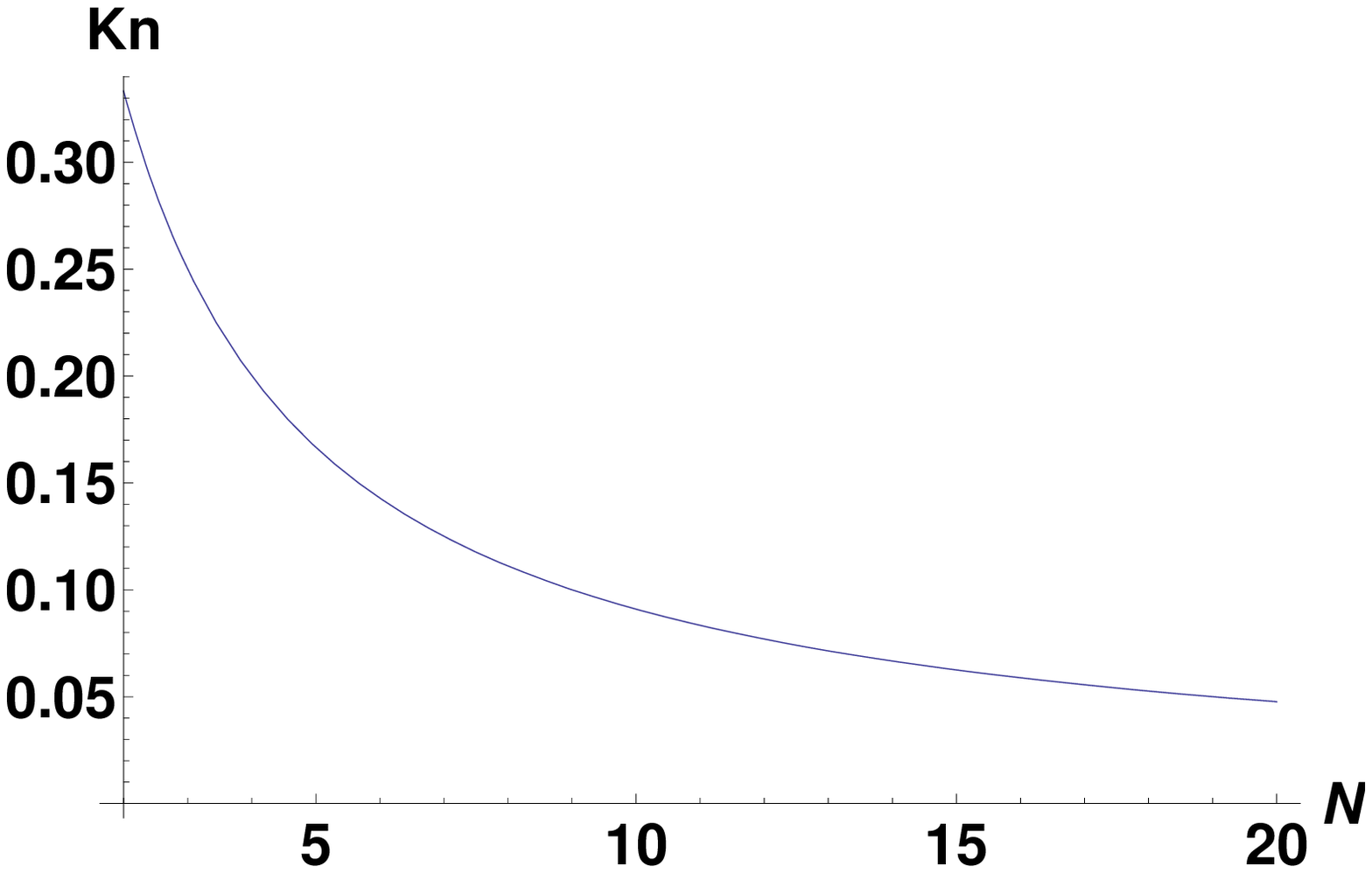 ,width=80mm}\\
\fig{Fig1}-a & \fig{Fig1}-b \\
\end{tabular}
\caption{\it The Debye's length $r_{D}$ value as function of the particles density $n$ at $T\,=\,160\,\,\,MeV$
in \fig{Fig1}-a and
the value of Knudsen number $Kn$ as function of $N$ from \eq{PartNumber} in \fig{Fig1}-b.}
\label{Fig1}
\end{figure}

Basing on the particles density $n\,\approx\,100\,\,\,fm^{-3}$ from \cite{Teaney} we can very roughly estimate the obtained value of the Knudsen number.
Indeed, the external potential is proportional to the overall charge of the clouds of the "fast particles"
which interact with the spot of "slow" particles:
\beq
\,V(Y)\,r_{0D}\,\propto\,q\,N(s)\,,
\eeq
where $N(s)$ is a number of particles in the cloud, it depends on the energy of the process. Therefore, very approximately we can write:
\beq\label{HydIneq1}
Kn\,\propto\,\,\frac{1}{\,1\,+\,N(s)}\,.
\eeq
The estimation of the Debye length at $T\,=\,160\,\,\,MeV$ gives $r_{0D}\,\approx\,0.15-0.3\,\,fm$ for different values
of $n$, see in \fig{Fig1}-a\footnote{Interesting to note, that this number is close to the size of a constituent quark obtained in \cite{Bond}.}. The number of the fast particles which interact with the hot spot we can estimate as 
\beq\label{PartNumber}
N\,\approx\,\pi\,r_{0D}^{2}\,R\,n\,,
\eeq
where $R$ is a size of the "fast" particles cloud in the longitudinal direction. 
The result of the calculations of $Kn$ from \eq{HydIneq1} as a function of the number of particles inside the
disk of the fast charged constituents is given in \fig{Fig1}-b. Of course, the value of $N$ crucially depends on the 
microscopic description of the interactions,  we can not determine $N$ precisely in our "toy" model.
Nevertheless, our results shows that we indeed traced the transition of the hot gas spot to the fluid state by the calculation of the change of the Debye radius of the spot.

\section{Conclusion}

In our paper, we considered a toy model of particles interacting at high energy and investigated 
the transition from gas to the liquid state of the system. The approach proposed has an analogy
in the inertial confinement effect in plasma physics and, additionally, is based on the following 
important propositions:
\begin{itemize}
\item  Reduction of the non-equilibrium process to the equilibrium one by the separation of the degrees of freedom.
The idea of \cite{Term} is that instead accounting of several degrees of freedom we can introduce an external field
which will interacts with other, different degrees of freedom.
\item The separation of the degrees of freedom in high-energy interactions. This is an old and well known idea, see
for example \cite{Eff1}. In our investigation, we separate the degrees of freedom based on Landau's paper \cite{Land1}.
There was assumed that at very early stages of interaction, a part of the interacting particles stopped and are in the rest.
The drops of these stopped particles is the main object of our investigation. We consider the properties of these drops
in the external field of the other, still relativistic particles. All relativistic dynamics holds
in the description of these "fast" particles and, therefore, in the description of the external field. The stopped particles are non-relativistic and can be described by usual techniques.
\item The creation of small charged drops of liquid matter in a whole scattering region is a local process. The parameters of the 
drops are varied locally, but all of them quickly achieve the state of equilibrium  due small sizes, which we assume to be proportional to the Debye's length.
\end{itemize}
Our main idea, thereby, is to trace the transition of the gas spots into the liquid phase in different models
based on these propositions.

For non-equilibrium system of interacting particles, it is almost impossible to write an equation of state
of the system in order to trace the gas-liquid transition. Instead, especially for complex systems,
this transition can be manifested by the change of so called Knudsen number, well known in hydrodynamics.
We calculated this number in different models, see \eq{Knud1}, \eq{Knud} and especially \eq{HydIneq}-\eq{HydIneq1}.
We see that in different models, our approach showed the liquid phase creation by the decrease
of the Knudsen number. The physical picture behind all these models is simple: like in plasma physics confinement effect, the external particles create an additional pressure on the initially stopped particles compressing dense gas
to the liquid state. The calculation of the Debye length changes, \eq{HydIneq}-\eq{HydIneq1}, is the most
important part of our investigation. For almost ideal fluid, the Vlasov's approximation to the Boltzmann equation 
found to be an appropriate one and similar description of the effects must take the place, to our opinion, in high energy nuclei-nuclei QCD scattering as well, 
see  \cite{PScr}.

Concerning the matter of the fluid's viscosity, we have to note papers  \cite{Skokov}, where
also the modified Van der Vaals equation was considered. There it was argued that
the viscosity effects are not small and play a significant role in the dynamics of the phase transition.
In general, it is difficult to compare our calculations with calculations of \cite{Skokov} because
we used the "toy" model only, in \cite{Skokov} much more realistic model was proposed and explored. 
Also, we consider the process of the drop compression whereas in \cite{Skokov} the system's expansion was mostly considered. Nevertheless,
we can underline an important fact that must be significant in  both models. 
In our calculations we consider a drops of small size, $r_{D}\,\approx\,0.2\,\,\,fm$ , as well as the authors of \cite{Skokov}. In this case the drop of the particles is not neutral anymore and it has a charge.
In the case of scattering of electron beams it is an electrical one, and, what is more important, it is a
color one in the case of the scattering of nuclei at high energy. Therefore, the subject of the consideration is a non-neutral plasma. The properties of this kind of plasma, including viscosity, are drastically different from the properties of the neutral one. Nor electric neither magnetic fields can not be neglected in the description of the dynamics
of charged plasma. Therefore, there is a question about an applicability of usual
Navier-Stokes equation in the situation when the framework similar to a magnetohydrodynamic approach
may be more appropriate.

The transport properties of the system under the 
external pressure also were considered 
in our model. For this purpose, we used the well known Enskog calculation scheme, see \eq{StateEq1} and \cite{Enskog,Prakash}. 
In this approach, we calculated the corrections to the transport coefficients
caused by the external influence on the gas whereas the initial values of the parameters are given, see
\eq{TrProp1}-\eq{TrProp3} and \eq{TrProp11}-\eq{TrProp33}.
Our results demonstrate that the transport properties of the system of interacting particles near the transition into the liquid phase
are similar, in some sense, to the properties of the quark-gluon plasma obtained in \cite{Rat}. Namely, we obtain similar  rise of the bulk viscosity when the temperature rises. The explanation of this effect in the model is simple. In Enskog approximation, the bulk viscosity is proportional to the density of the state. The density grows toward some constant value whereas the system of interest undergoes the gas-liquid transition. When the state of ideal liquid is created then the density is constant due to incompressibility of the state and, therefore, the bulk viscosity as well.

  There are also the following important issues arose in our framework which were designated but not fully investigated in the manuscript. The drop of "slow", stopped particles has a non zero charge. Definitely it must affect on the transport properties of the drop, which is located in the external magnetic and electric fields, and it must affect on the transport properties of whole bulk of scattering matter. The evolution of this mixture of liquid drops and dense quark-gluon gas is an another difficult problem, see also in \cite{Skokov}. We assumed, that the characteristic of the drops are varied from point to point, therefore in the system some long-range interactions must be present
\footnote{Long-range interactions are needed in the light of the Mermin-Wagner theorem.}. In some sense, in our model we have a non-equilibrium system with equilibrated drops of matter inside. Perhaps the statistical description of this system is possible in the framework of so-called "super statistical" description, see\cite{SuperStat}.

 Another issue is a matter of strong correlation between the particles in the QGP, see \cite{Shyr1}.
It is well known, see \cite{CorrJapan}, that
strong correlation means the small number of the particles in the volume determined by the Debye's length.
In turn, this contradicts to
the assumption of the applicability of the Vlasov's equation and to the assumption of the
thermodynamical equilibrium of the matter in the interaction area.
The possible solution of this contradiction may be found in a small size of the created liquid spots, the  thermalization of them happens very fast and it is local. Perhaps, smallness of the drops will allow to describe
strongly correlated equilibrated and charged particles inside the drops of hot matter. We note,
that in our calculations the number of the particles inside the volume of the spot is indeed small because of
the smallness of the Debye's length.

 The question about the magnetic field and additional degrees of freedom in the drop's creation and evolution
was not considered in the paper, in spite of the fact that it is very important problem, see \cite{NNPlasma}.
Indeed, charged drops created in high energy interaction will rotate. The  subsequent drop's evolution
dynamics, therefore, is pretty complicated, it will include also additional possible instability effects
in the system of charged particles under the external pressure. All these together will influence on multiplicities of particles produced in high energy interactions. The discovering of the traces of these complicated dynamics additionally to the traces of the phase transition in the particle's production experiments is a very interesting task for the future studying.

  Finally we conclude that the main purpose in developing of our toy model was to designate some basic principles and methods for the calculations of the gas-liquid transition in the system of relativistically interacting particles.
More detailed analysis
of the hot drops condensation in the framework of QED and QCD is the aim of our future work, see \cite{Prep}.

%%%%%%%%%%%%%%%%%%%%%%%%%%%%%%%%%%%%%%%%%%%%%%%%%%%%%%%%%%%%%%%%%%%%%%%%%%%%

\newpage
\section*{Appendix A: gas of interacting particles}

\renewcommand{\theequation}{A.\arabic{equation}}
\setcounter{equation}{0}

In this appendix we shortly remind the main facts
concerning a virial expansion, see the detailed derivation in \cite{Land}. 
We consider a gas of N interacting particles each with mass $m$ on a plane as a gas of  hard disk with small 
thickness and radius $r_0$. The energy of the gas in the classical limit is given by the well known expression:

\beq\label{TEn}
E(p,q)=\sum^{N}_{i=1}\frac{p_{i}^2}{2\,m}\,+\,U(b_{1},...,b_{N})\,
\eeq
where as usual the first term is the kinetic energy of N particles, $U$ is a potential energy
of their mutual interactions and $b_{1},...,b_{N}$ their coordinates. The grand partition function for this Hamiltonian is
\beq
Q(\mu,T,S)\,=\,e^{-\beta\,\Omega}\,=\,\sum_{N=0}^{\infty}\,e^{\beta\,\mu\,N}\,\frac{1}{N!}\,
\int\,\prod_{i=1}^{N}\,\frac{d^{2}\vec{p}_{i}\,d^{2}\vec{b}_{i}}{h^2}\,\exp(-\beta\,\sum^{N}_{i=1}\frac{p_{i}^2}{2\,m}\,)\,
\,\exp(-\beta\,\,U(b_{1},...,b_{N}))\,.
\eeq
where $\Omega$ is the grand potential of the problem. As usual, integrating over momenta, we obtain
\beq
Q(\mu,T,S)\,=\,e^{-\beta\,\Omega}\,=\,\,\sum_{N=0}^{\infty}\,\left(\frac{e^{\beta\,\mu}}{\lambda^{2}}\right)^{N}\,
\frac{Z_{N}(S,\beta)}{N!}
\eeq
and
\beq
Z_{N}(S,\beta)\,=\,\int...\int\,d^{2}\vec{b}_{1},...,d^{2}\vec{b}_{N}\,\exp\left(-\beta\,\,U(b_{1},...,b_{N})\right)\,.
\eeq
Here $\lambda$ is the De Broglie length of the quark corresponding to the average energy $\beta$:
\beq
\lambda\,=\,\left(\frac{h^2\,\beta}{2\,\pi\,m}\right)^{1/2}\,.
\eeq 
The potential of the problem is given by
\beq
\Omega\,=\,-\frac{1}{\beta}\,\ln\left(\,1+\frac{e^{\beta\,\mu}}{\lambda^{2}}\,S\,+\,
\frac{e^{2\,\beta\,\mu}}{2!\,\lambda^4}\,\int\int\,d^{2}\vec{b}_{1}\,d^{2}\vec{b}_{2}\,
\exp\left(-\beta\,\,U(b_{1},b_{2})\,\right)\,+ ...\Ra\,.
\eeq 
Here, we used 
\beq
\int\,\,d^{2}\vec{b}\,=\,S\,=\pi\,R^{2}\,,
\eeq
where $R^{2}$ is the characteristic radius of the problem.
In the following we will define and consider only pairwise interaction between the particles, namely we have
\beq
\,U(b_{1},b_{2})\,=\,U(|b_{1}-b_{2}|)\,=\,U(b_{12})\,=\,U(b)\,=\,U_{12}\,
\eeq
and therefore, in the relative  coordinates of the center mass 
we reduce the multiplicity of the integrated functions and obtain an additional $S$ factor in the integrals:
\beq
\Omega\,=\,-P\,S\,=-\frac{1}{\beta}\,\ln\left(\,1\,+\,S\,\frac{e^{\beta\,\mu}}{\lambda^{2}}\,+\,
S\,\frac{e^{2\,\beta\,\mu}}{2!\,\lambda^4}\,\int\int\,d^{2}\vec{b}_{12}\,
\exp\left(-\beta\,\,U_{12}\,\right)\,+ ...\Ra\,.
\eeq 
Introducing variable $\zeta$
\beq
\zeta\,=\,\frac{e^{\beta\,\mu}}{\lambda^2}\,
\eeq
we obtain the expression for the potential in the form of the series in $\zeta$
\beq\label{Press}
\Omega\,=\,-P\,S\,=-\frac{S}{\beta}\,\sum_{n=1}^{\infty}\,\frac{J_{n}}{n!}\,\zeta^{n}\,.
\eeq
We will take into account only two first terms of this series with the following $J_1$ and $J_2$: 
\beq\label{Integr}
J_1\,=\,1\,,\,\,\,\,J_2\,=\,\int\int\,d^{2}\vec{b}_{12}\,\Le\,
\exp\left(-\beta\,\,U_{12}\,\right)\,-\,1\,\Ra.
\eeq
The number of particles in this gas we obtain as usual
\beq
N=-\left(\frac{\partial\,\Omega}{\partial\,\mu}\Ra_{T,S}
\eeq
and because $\partial\,\zeta\,/\,\partial\,\mu\,=\,\beta\,\zeta\,$ we finally have:
\beq\label{PNumb}
N\,=\,S\,\sum_{n=1}^{\infty}\,\frac{J_{n}}{(n-1)!}\,\zeta^{n}\,.
\eeq
Excluding from \eq{Press} and \eq{PNumb} the variable $\zeta$, we obtain in the second order approximation
the equation of state for our gas:
\beq
P\,=\,\frac{NT}{S}\,-\,\frac{N^2\,T}{2\,S^2}\,J_{2}\,,
\eeq
where $T=1/\beta$.

\newpage
\section*{Appendix B: deviation from equilibrium state and non static electrical field of the gas in Vlasov's approximation }

\renewcommand{\theequation}{B.\arabic{equation}}
\setcounter{equation}{0}

In the linear approximation over $f_{0}(r,p)\,$, we have the Boltzmann equation
\beq\label{Bolz11}
\frac{\partial\,f_{1}(r,p,t)}{\partial\,t}\,+\,v\,\frac{\partial\,f_{1}(r,p,t)}{\partial\,r}\,+\,
\Le\,-\,\frac{\partial\,V_{2}(r)}{\partial\,r}\,+\,q\,E^{0}\,\Ra\,\frac{\partial\,f_{1}(r,p,t)}{\partial\,p}\,+\,
q\,E^{1}\,\frac{\partial\,f_{0}(r,p)}{\partial\,p}\,=\,0\,
\eeq
together with the Maxwell's equations for the electric field:
\beq\label{Max11}
rot\,E^{1}\,=\,0\,,\,\,\,\,div\,E^{1}\,=\,4\,\pi\,q\,n\,\int\,f_{1}(r,p,t)\,d^{3}\,p\,,
\eeq
where
\beq
E^{0}\,=\,-\,grad\,\phi_{0}(r)
\eeq
with $\phi_{0}(r)$ given by \eq{Equil}. Solutions for the $f_{1}(r,p,t)$ distribution function and 
$E^{1}$ field we are deriving below. 

Proceeding with the equation \eq{Bolz1} we perform a following substitution:
\beq\label{Dev1}
f_{1}(r,p,t)\,=\,\bar{f}_{1}(r,p,t)\,f_{0}(r,p)
\eeq
with $f_{0}(r,p)$ from \eq{Equil}. Rewriting \eq{Bolz11} we obtain:
\beq\label{Bolz22}
\frac{\partial\,\bar{f}_{1}(r,p,t)}{\partial\,t}\,+\,
\frac{p}{m}\,\frac{\partial\,\bar{f}_{1}(r,p,t)}{\partial\,r}\,+\,
F\,\frac{\partial\,\bar{f}_{1}(r,p,t)}{\partial\,p}\,-\,
\frac{q\,E^{1}}{m\,k_{B}\,T\,}\,p\,=\,0\,,
\eeq
where $F$ is given by \eq{Force}. We consider a linear approximation over the equilibrium
distribution, therefore, we could write the external force in the equation in following form
\beq
F(r)\,\rightarrow\,\bar{F}\,=\,F(\bar{r})\,=\,
-\,\Le\,\frac{\partial\,V_{2}(r)}{\partial\,r}\,\Ra_{r\,=\,\bar{r}}\,+\,q\,E^{0}(\bar{r})\,.
\eeq
with $r$  
\beq
\bar{r}\,=\,\frac{\int\,r\,f_{0}(r,p)\,d^{3}x\,d^{3}p\,}{\,\int\,f_{0}(r,p)\,d^{3}x\,d^{3}p}\,.
\eeq
Thereby we have the following equation for the $\bar{f}_{1}(r,p,t)$ distribution function:
\beq\label{Bolz222}
\frac{\partial\,\bar{f}_{1}(r,p,t)}{\partial\,t}\,+\,
\frac{p}{m}\,\frac{\partial\,\bar{f}_{1}(r,p,t)}{\partial\,r}\,+\,
\bar{F}\,\frac{\partial\,\bar{f}_{1}(r,p,t)}{\partial\,p}\,-\,
\frac{q\,E^{1}}{m\,k_{B}\,T\,}\,p\,=\,0\,
\eeq
Performing Fourier transform of $\bar{f}_{1}(r,p,t)$ and $E^{1}(r,t)$
\beq\label{Four1}
\bar{f}_{1}(r,p,t)\,=\,\int\frac{d\omega}{2\,\pi}\int\frac{d^{3}k}{(2\,\pi)^3}\,
e^{-i\omega\,t\,+\,i\,r\,k}\,\phi_{0}(k,p,\omega)
\eeq 
\beq
E^{1}(r,t)\,=\,\int\frac{d\omega}{2\,\pi}\int\frac{d^{3}k}{(2\,\pi)^3}\,
e^{-i\omega\,t\,+\,i\,r\,k}\,\tilde{E}^{1}(k,\omega)
\eeq
we obtain finally a nonlinear differential equation of the first order over momenta:
\beq\label{Bolz3}
\frac{\partial\,\phi_{0}}{\partial\,p}\,+\,a\,p\,\phi_{0}\,-\,b\,\phi_{0}\,-\,c\,p\,=\,0\,.
\eeq
Here
\beq
a\,=\,\frac{ik}{m\,\bar{F}}\,,\,\,\,b\,=\,\frac{i\,\omega}{\bar{F}}\,,\,\,\,
c\,=\,\frac{q\,\tilde{E}^{1}}{\,m\,k_{B}\,T\,\bar{F}}\,.
\eeq
The solution of this equation, with additional condition that at $p\,\rightarrow\,0\,(T\,\rightarrow\,0\,\,)$
the solution is static, i.e. $\phi_{0}(k,p\,=\,0\,,\omega)\,=\,0\,$,  is the following function:
\beq
\phi_{0}(k,p,\omega)\,=\,c\,e^{-a\,p^2\,/\,2\,+\,b\,p}\,\int_{0}^{p}\,p^{'}\,
e^{a\,p^{'2}\,/\,2\,-\,b\,p^{'}}\,dp^{'}\,.
\eeq
Thereby we obtain for the correction to the equilibrium state:
\beq
\phi_{0}(k,p,\omega)\,=\,\frac{q\,\tilde{E}^{1}}{\,m\,k_{B}\,T\,\bar{F}}\,
e^{-a\,p^2\,/\,2\,+\,b\,p}\,\int_{0}^{p}\,p^{'}\,
e^{a\,p^{'2}\,/\,2\,-\,b\,p^{'}}\,dp^{'}\,.
\eeq
We see, that this correction is suppressed in comparison to the distribution
function \eq{Equil} by the factor $\frac{q\,\tilde{E}^{1}}{\bar{F}}$.
Substituting this expression back in \eq{Four1}, we have:
\beq
\bar{f}_{1}(r,p,t)\,=\,\frac{q}{\,m\,k_{B}\,T\,\bar{F}}\,
\int_{0}^{p}\,dp^{'}\,p^{'}\,
\int\frac{d\omega}{2\,\pi}\int\frac{d^{3}k}{(2\,\pi)^3}\,\tilde{E}^{1}(k,\omega)\,
e^{-i\omega\,\Le\,t\,+\,\frac{p^{'}}{\bar{F}}\,-\,\frac{p}{\bar{F}}\,\Ra\,
+\,i\,k\,\Le\,r\,+\,\frac{p^{'2}}{2\,m\,\bar{F}}\,-\,\,\frac{p^{2}}{2\,m\,\bar{F}}\,\Ra}\,.
\eeq
Performing Fourier transform again, we obtain for our function in $(r,t)$ representation:
\beq\label{NonEqD}
\bar{f}_{1}(r,p,t)\,=\,\frac{q}{\,m\,k_{B}\,T\,\bar{F}}\,\int_{0}^{p}\,dp^{'}\,p^{'}\,
E^{1}(r\,+\,\frac{p^{'2}\,-\,p^{2}}{2\,m\,\bar{F}}\,,
t\,+\,\frac{p^{'}\,-\,p\,}{\bar{F}}\,)\,.
\eeq
This correction to the static distribution function determines also the non-static electric field which we consider further.

Using \eq{Max}, we find the equation for the correction $E^{1}$ to the electric field:
\beq\label{Dev2}
div\,E^{1}(r,t)\,=\,\frac{4\,\pi\,q^2\,n}{\,m\,k_{B}\,T\,\bar{F}}\,
\int\,d^{3}\,p\,\int_{0}^{p}\,dp^{'}\,p^{'}\,E^{1}(r\,+\,\frac{p^{'2}\,-\,p^{2}}{2\,m\,\bar{F}}\,,
t\,+\,\frac{p^{'}\,-\,p\,}{\bar{F}}\,)\,f_{0}(r,p)\,
\eeq
with with $f_{0}(r,p)$ from \eq{Equil}, see also \eq{Dev1}. This equation is highly non-linear and non-local and,
perhaps, a precise solution of the equation can be found only by numerical methods. In order to investigate  approximate solutions
of the equation, we perform Fourier transform of the functions in \eq{Dev2}:
\beq
k\tilde{E}^{1}(k,\omega)=\frac{4\pi\,q^2\,n}{m\,k_{B}\,T\bar{F}}
\int d^{3}p\int_{0}^{p} dp^{'}p^{'}\int \frac{d^{3}k^{'}}{(2\pi)^{3}}
\tilde{E}^{1}(k-k^{'},\omega)e^{-i\omega\Le\frac{p^{'}-p}{\bar{F}}\Ra
+i(k-k^{'})\Le\frac{p^{'2}-p^{2}}{2m\bar{F}}\,\Ra}\,
\tilde{f}_{0}(k^{'},p)\,,
\eeq
where
\beq
f_{0}(r,p)\,=\,\int\frac{d^{3}k}{(2\,\pi)^3}\,
e^{\,i\,r\,k}\,\tilde{f}_{0}(k,p)\,.
\eeq
In the right hand side of the equation the oscillating integral over $k^{'}$ is not vanishing when
\beq
k^{'}\,\propto\,\frac{2m\bar{F}}{p^{'2}-p^{2}}\,\propto\,\frac{\bar{F}}{k_{B}\,T}\,\ll\,1\,
\eeq
in our approximation of the weak external field.
Therefore, in the region where
\beq\label{KCond}
k\,>\,k^{'}\,
\eeq
in the first approximation over $k^{'}$ we have:
$$
k\tilde{E}^{1}(k,\omega)\approx\tilde{E}^{1}(k,\omega)
\frac{4\pi\,q^2\,n}{m\,k_{B}\,T\bar{F}}
\int d^{3}p\int_{0}^{p} dp^{'}p^{'}\int \frac{d^{3}k^{'}}{(2\pi)^{3}}
e^{-i\omega\Le\frac{p^{'}-p}{\bar{F}}\Ra
+i(k-k^{'})\Le\frac{p^{'2}-p^{2}}{2m\bar{F}}\,\Ra}\,
\tilde{f}_{0}(k^{'},p)\,-\,
$$
\beq\label{Dev4}
\,-\,\frac{\partial\,\tilde{E}^{1}(k,\omega)}{\partial\,k}\,
\frac{4\pi\,q^2\,n}{m\,k_{B}\,T\bar{F}}
\int d^{3}p\int_{0}^{p} dp^{'}p^{'}\int \frac{d^{3}k^{'}}{(2\pi)^{3}}\,k^{'}\,
e^{-i\omega\Le\frac{p^{'}-p}{\bar{F}}\Ra
+i(k-k^{'})\Le\frac{p^{'2}-p^{2}}{2m\bar{F}}\,\Ra}\,
\tilde{f}_{0}(k^{'},p)\,
\eeq
or, simplifying the notations we obtain:
\beq
\frac{\partial\,\tilde{E}^{1}(k,\omega)}{\partial\,k}\,-\,\tilde{E}^{1}(k,\omega)\,\Le\,
C_{1}(\omega,k)\,-\,C_{2}(\omega,k)\,
k\,\Ra\,=\,0\,,
\eeq
where
\beq
C_{1}(\omega,k)\,=\,\frac{\int d^{3}p\int_{0}^{p} dp^{'}p^{'}\,e^{-i\omega\Le\frac{p^{'}-p}{\bar{F}}\Ra
+i\,k\Le\frac{p^{'2}-p^{2}}{2m\bar{F}}\,\Ra}\,
f_{0}(\frac{p^{2}-p^{'2}}{2m\bar{F}},p)\,}
{\int d^{3}p\int_{0}^{p} dp^{'}p^{'}\int \frac{d^{3}k^{'}}{(2\pi)^{3}}\,k^{'}\,
e^{-i\omega\Le\frac{p^{'}-p}{\bar{F}}\Ra
+i\,(k-k^{'})\Le\frac{p^{'2}-p^{2}}{2m\bar{F}}\,\Ra}\,
\tilde{f}_{0}(k^{'},p)\,}
\eeq
and
\beq
C_{2}(\omega,\,k)\,=\,\frac{m\,k_{B}\,T\bar{F}}
{4\pi\,q^2\,n\,\int d^{3}p\int_{0}^{p} dp^{'}p^{'}\int \frac{d^{3}k^{'}}{(2\pi)^{3}}\,k^{'}\,
e^{-i\omega\Le\frac{p^{'}-p}{\bar{F}}\Ra
+i\,(k-k^{'})\Le\frac{p^{'2}-p^{2}}{2m\bar{F}}\,\Ra}\,
\tilde{f}_{0}(k^{'},p)\,}\,.
\eeq
An integration of \eq{Dev4} gives
\beq
\tilde{E}^{1}(k,\omega)\,=\,
\,C_{0}(\omega)\,e^{\,\int^{k}_{0}\,C_{1}(\omega,\,t)\,dt\,-\int^{k}_{0}\,C_{2}(\omega,\,t)\,t\,dt}\,.
\eeq
We see, that the fluctuation of electric field in this region of $k$ is suppressed  by the large
factor $C_{2}$ in the power of exponent.

In the opposite limit, when
\beq
k\,\sim\,k^{'}\,\propto\,\frac{2m\bar{F}}{p^{'2}-p^{2}}\,\propto\,\frac{\bar{F}}{k_{B}\,T}\,\ll\,1\,
\eeq
we introduce new variable $\epsilon$:
$$
\epsilon\,=\,k\,-\,k^{'}.
$$
Equation \eq{Dev4} will acquire the following form:
\beq\label{Dev5}
k\tilde{E}^{1}(k,\omega)=\,\frac{4\pi\,q^2\,n}{m\,k_{B}\,T\bar{F}}
\int d^{3}p\int_{0}^{p} dp^{'}p^{'}\int \frac{d^{3}\epsilon}{(2\pi)^{3}}
\tilde{E}^{1}(\epsilon,\omega)e^{-i\omega\Le\frac{p^{'}-p}{\bar{F}}\Ra
+i\epsilon\Le\frac{p^{'2}-p^{2}}{2m\bar{F}}\,\Ra}\,
\tilde{f}_{0}(k-\epsilon,p)\,.
\eeq
Fourier transform of function \eq{Equil} is
\beq
\tilde{f}_{0}(k,p)=\tilde{f}_{0}(k)f_{0}(p)=f_{0}(p)
\int\,d^{3}r\,
e^{-i r\,k}e^{-\frac{V_{2}(r)}{k_{B}T}-\frac{q\phi_{0}(r)}{k_{B}T}}\approx
f_{0}(p)\Le
(2\pi)^{3}\delta^{3}(k)-\frac{\tilde{V}_{2}(k)}{k_{B}T}-\frac{q\tilde{\phi}_{0}(k)}{k_{B}T}\Ra
\eeq
with
\beq
f_{0}(p)\,=\,\frac{1}{\Le\,2\,\pi\,k_{B}\,T\,\Ra^{3/2}}\,
e^{-\frac{p^2}{2mk_{B}T}\,}\,,
\eeq
therefore, in the first approximation of expansion over $\epsilon$ we obtain:
$$
\tilde{E}^{1}(k,\omega)=\,\tilde{E}^{1}(k,\omega)\,\frac{4\pi\,q^2\,n}{m\,k_{B}\,T\bar{F}\,k}
\int d^{3}p\,f_{0}(p)\,\int_{0}^{p} dp^{'}p^{'}\,e^{-i\omega\Le\frac{p^{'}-p}{\bar{F}}\Ra
+i\,k\Le\frac{p^{'2}-p^{2}}{2m\bar{F}}\,\Ra}\,-\,
$$
$$
\,-\,\tilde{E}^{1}(0,\omega)\,\frac{4\pi\,q^2\,n\,\tilde{V}_{2}(k)}{m\,(k_{B}\,T)^2\,\bar{F}\,k}
\int d^{3}p\,f_{0}(p)\,\int_{0}^{p} dp^{'}p^{'}\,e^{-i\omega\Le\frac{p^{'}-p}{\bar{F}}\Ra}\,
\int^{\frac{\bar{F}}{k_{B}\,T}\,} \frac{d\epsilon\,\epsilon^{2}}{2\pi^{2}}\,-\,
$$
\beq
\,-\,\,\tilde{E}^{1}(0,\omega)\,\frac{4\pi\,q^3\,n\,\tilde{\phi}_{0}(k)}{m\,(k_{B}\,T)^2\,\bar{F}\,k}
\int d^{3}p\,f_{0}(p)\,\int_{0}^{p} dp^{'}p^{'}\,e^{-i\omega\Le\frac{p^{'}-p}{\bar{F}}\Ra}\,
\int^{\frac{\bar{F}}{k_{B}\,T}\,} \frac{d\epsilon\,\epsilon^{2}}{2\pi^{2}}\,.
\eeq
Finally we obtain for our electric field:
\beq
\tilde{E}^{1}(k,\omega)\,\varepsilon\,=\,-\,\tilde{E}^{1}(0,\omega)\,
\frac{2\,q^2\,n\,\bar{F}^{2}\,}{3\,\pi\,m\,(k_{B}\,T)^{4}\,\,k}\,
\Le\,\frac{\tilde{V}_{2}(k)}{k_{B}T}+\frac{q\tilde{\phi}_{0}(k)}{k_{B}T}\Ra
\int d^{3}p\,f_{0}(p)\,\int_{0}^{p} dp^{'}p^{'}\,e^{-i\omega\Le\frac{p^{'}-p}{\bar{F}}\Ra}\,
\eeq
where
\beq\label{Diel}
\varepsilon\,=\,1\,-\,\frac{4\pi\,q^2\,n}{m\,k_{B}\,T\bar{F}\,k}
\int d^{3}p\,f_{0}(p)\,\int_{0}^{p} dp^{'}p^{'}\,e^{-i\omega\Le\frac{p^{'}-p}{\bar{F}}\Ra
+i\,k\Le\frac{p^{'2}-p^{2}}{2m\bar{F}}\,\Ra}\,
\eeq
is the dielectric constant of the problem.

%%%%%%%%%%%%%%%%%%%%%%%%%%%%%%%%%%%%%%%%%%

%%%%%%%%%%%%%%%%%%%%%%%%%%%%%%%%%%%%%%%%%%

\end{document}